\documentclass[final,5p,times,twocolumn]{elsarticle}

\usepackage{amssymb}

\usepackage{color}

\usepackage{amsmath}

\journal{Computer Physics Communication}

\begin{document}

\begin{frontmatter}



\title{ASCOT: solving the kinetic equation of\\ minority particle species in tokamak plasmas}


\author{E. Hirvijoki\corref{cor1}}
\cortext[cor1]{Corresponding author.}
\ead{eero.hirvijoki@aalto.fi}
\author{O. Asunta}
\author{T. Koskela}
\author{T. Kurki-Suonio}
\author{J. Miettunen}
\author{S. Sipil\"{a}}
\author{A. Snicker}
\author{S. \"{A}k\"{a}slompolo}

\address{Aalto University, Department of Applied Physics, P.O. Box 14100, FI-00076 AALTO, Finland}

\begin{abstract}
A comprehensive description of methods, suitable for solving the kinetic equation for fast ions and impurity species in tokamak plasmas using Monte Carlo approach, is presented. The described methods include Hamiltonian orbit-following in particle and guiding center phase space, test particle or guiding center solution of the kinetic equation applying stochastic differential equations in the presence of Coulomb collisions, neoclassical tearing modes and Alfv\'en eigenmodes as electromagnetic perturbations relevant to fast ions, together with plasma flow and atomic reactions relevant to impurity studies. Applying the methods, a complete reimplementation of the well-established minority species code ASCOT is carried out as a response both to the increase in computing power during the last twenty years and to the weakly structured growth of the code, which has made implementation of additional models impractical. Also, a benchmark between the previous code and the reimplementation is accomplished, showing good agreement between the codes.
\end{abstract}


\end{frontmatter}


\section{Introduction}
\label{sec:intro}
As fusion energy research is rapidly approaching the reactor era, the role of energetic ions, most notably fusion alphas, is becoming increasingly important. Numerical simulation tools for analyzing their behavior have been available for over 20 years. During this period, the demands of more accurate and complete physics on one hand, and the development of computer architectures on the other hand have forced the developers to alter and amend the codes piece by piece, easily leading to a situation where the original philosophy and structure of the code are compromised and the full command of it is lost as a result of what could be called organic growth. The Monte Carlo orbit-following code ASCOT~\cite{ascot1995,kiviniemi:cjp:2001,ascot_wall2009,kurki-suonio2011,asuntaSimsAUGRMP,ErNBIionsatAUG} is no exception in this.

In this paper we introduce a code that is still called ASCOT but has been fully rewritten. The new version is dubbed ASCOT4 and, when necessary, the old version is referred to as ASCOT3. Not only is the code rewritten using modern programming conventions but, more importantly, the philosophy behind the model has been reviewed. When doing so it was realized that the conventional way of solving the guiding center kinetic equation including collisional effects is in fact inconsistent: guiding center codes typically apply a Monte Carlo collision operator that has not been transformed to the guiding center phase space and, strictly speaking, is valid only for the particle phase space. Another common discrepancy is to use different phase space coordinates for the Hamiltonian and collisional parts when solving the kinetic equation with Monte Carlo methods. In ASCOT4, we follow recent developments regarding these issues and treat the collisional and Hamiltonian parts consistently~\cite{brizard:fokkerPlanck,hirvijoki:092505:2013:pop}.

The paper is organized as follows: in Section~\ref{sec:ASCOT}, an introduction to ASCOT as a test particle code is given. Section~\ref{sec:equations} forms the backbone of this paper, describing in detail how the kinetic equation corresponding to the charged particle dynamics in tokamak plasmas is solved both for the full gyro motion and in the guiding center formalism. In this context we present the equations of motion for charged particles in electromagnetic fields, together with the integrators for advancing these equations in time, as well as methods to account for test particle Coulomb collisions with the bulk plasma. The initial particle loadings corresponding to the most relevant fast ion sources are described in Section~\ref{sec:sources}. In Section~\ref{sec:mhd}, we discuss how magnetohydrodynamical (MHD) effects can be included to obtain more realistic simulations of fast ions, and Section~\ref{sec:impurity} introduces models for ionization and recombination required in impurity studies. Section~\ref{sec:wall} discusses the implementation of a wall surrounding the tokamak plasma, i.e., the first solid surface of the machine seen by the particles, and how it limits the simulation regime. The rest of the paper includes a description of distributions that can be recorded during the simulation (Sec.~\ref{sec:diagnostics}), a benchmark between the new and the old version of the code (Sec.~\ref{sec:benchmark}), and a section describing various features of the code related to different platforms (Sec.~\ref{sec:hightech}), as well as a list of libraries required for compiling and executing the code on different platforms.

\section{ASCOT -- Monte Carlo code for minority species in tokamak plasmas}
\label{sec:ASCOT}
The ASCOT code was originally designed for accelerated simulation of charged particle orbits in tokamaks, and the first targeted studies concerned runaway electrons as well as fast ion current drive in simple axisymmetric magnetic backgrounds~\cite{Heikkinen1993215}. Early on ASCOT was upgraded to operate with realistic magnetic backgrounds that allowed shaped plasmas, up-down asymmetries as well as an X-point and a scrape-off layer (SOL). The toroidal non-uniformity included only the ripple produced by the finite number of toroidal field (TF) coils, and was given by trigonometric functions, sine or cosine, multiplied by an experimentally measured radial profile of the ripple strength. With the inclusion of a radial electric field model, ASCOT was used to model the dynamics of NBI ions at the plasma edge in the presence of both collisions and a radial electric field~\cite{kurkisuonio:CTPP:CTPP2150340214}. In particular, the response of ions trapped toroidally (i.e., between two adjacent TF coils) to the appearance/disappearance of the edge radial electric field was of interest~\cite{heikkinen:3655,Heikkinen:nucl:fusion:38:1998,heikkinen:phys:plasmas:692:1998,heikkinen:ppcf:679:1998,herrmann:ppcf:683:1998,activeEr,Heidbrink:ppcf:2001:373}.

Due to the increasing size of the simulations and its practically ideal multiprocessor scalability, ASCOT was parallelized using MPI in the late 1990's. This made possible much larger simulations with several hundreds and even thousands of parallel processes. When the polarization equation was included in ASCOT, even bulk ions could be simulated, and the evolution of the radial electric field due to non-ambipolar currents was calculated in conditions characteristic of a low to high (L-H) confinement mode transition in the ASDEX Upgrade tokamak~\cite{heikkinen:ppcf:693:1998,kiviniemi:chech:j:phys:1999,kiviniemi:ppcf:A185:2000,heikkinen:pop:2824:2001,Heikkinen2001527,kiviniemi:cjp:2001}. The formation of a transport barrier via this mechanism was also investigated for the small FT-2 tokamak at Ioffe Institute, St. Petersburg~\cite{lashkul,kurkisuonio:ppcf:44:301:2002,kurki-suonio:072510}. The significance of ion orbit losses to the divertor power loads was studied at JET~\cite{Fundamenski2003787,Matthews2003986} and, since ASCOT allowed simulations also in the SOL, it was used to investigate the divertor in/out asymmetries observed at JET~\cite{kurkisuonio:cjp:2001,kurkisuonio:nucl:fusion:42:725:2002,fundamenski:ppcf:44:761:2002}. For the same reason, ASCOT was also applied to kinetic electrons in the ASDEX Upgrade SOL, trying to identify reasons for the discrepancy between results from the SOLPS code and measurements~\cite{ahomantila:ppcf:50:065021:2008}.

For nearly ten years now, ASCOT has been used almost solely to simulate energetic ions. About ten years ago ASCOT also reached full maturity as far as the magnetic backgrounds are concerned: an arbitrary 3D field can be utilized, facilitating simulations in the presence of non-periodic features such as the test blanket modules (TBM) in ITER. At the same time the wall collision model was also upgraded to 3D: a wall surface consisting of triangular and quadrilateral elements was introduced as a  limit of the simulation regime. Since then, ASCOT has been predominantly used in 3D configuration. These studies include calculations of fast ion power loads on ITER first wall components, including contributions from fusion alphas, NBI ions and ICRH-generated ions~\cite{ascot_wall2009,kurki-suonio2011}, simulations of first wall power loads on ASDEX Upgrade in the presence of the then-new ELM mitigation coils~\cite{asuntaSimsAUGRMP}, and simulations of the NBI losses in the TBM mock-up experiments at DIII-D~\cite{kramer:nf:51:103029}. In this context, changes in the neutron production were also evaluated and compared to measurements with very good agreement. This required simulations of fusion-born tritium in DIII-D, which were made possible by the recent ASCOT code enhancement that allows following the full gyro orbits instead of guiding center orbits~\cite{snicker2010,Snicker2012_fusion_alpha_ITER}.

The most recent application for ASCOT is found at the low end of the energy spectrum: impurity transport, important for tritium retention, material migration and plasma performance in fusion reactors, is generally studied using axisymmetric codes that have a limited computational domain. ASCOT, for its part, offers a 3D alternative with the ability to follow particles in an unrestricted domain ranging from the core plasma to the first wall. For modeling impurity transport, ASCOT was enhanced to include relevant atomic physics and a background plasma flow pertinent to the SOL region. A trace-element injection experiments carried out at ASDEX Upgrade were then simulated using the code. The results revealed that, in contrast to the commonly used experimental assumption of pure toroidal symmetry, the impurity deposition pattern can exhibit a strong toroidal asymmetry due to the 3D features of the first wall~\cite{miettunen:NF2012}. Since then, similar modeling has been carried out also for JET to aid the interpretation of a beryllium migration experiment~\cite{miettunen:JNM2013}.

As is evident from the above history, the code has grown tremendously since the early 1990's as new methods and models have been incorporated. The work has been carried out by a number of developers, using a range of programming conventions. Simultaneously, the vast increase in computing resources with new supercomputers available around the world has forced the code out of the serial and modestly parallel processing era to accommodate up to tens of thousands of parallel processes. 

\section{Kinetic equation for minority particle species}
\label{sec:equations}
ASCOT is often used to calculate quantities such as fast ion density and torques for other codes to use as input. Thus, solving the distribution function of the minority species forms the backbone of the code. The time evolution of the distribution function $f(\mathbf{z},t)$ of an ensemble of test particles in a plasma is described by the kinetic equation
\begin{align}
\frac{\partial f}{\partial t}+\mathbf{\dot{z}}\cdot\frac{\partial f}{\partial\mathbf{z}}=\left(\frac{\partial f}{\partial t}\right)_{coll},
\label{eq:motion_fokker_planck}
\end{align}
where $\mathbf{z}=(\mathbf{r},\mathbf{v})$ is the particle phase space, $\mathbf{\dot{z}}$ stands for the equations of motion, and $(\partial f/\partial t)_{coll}$ describes the change in $f$ due to collisional processes. As the collisional effects, the right-hand side of Eq.~(\ref{eq:motion_fokker_planck}), are often modelled with diffusion and friction, the kinetic equation essentially becomes a six-dimensional partial differential equation.

In practice, the high dimensionality excludes finite element and finite difference methods for finding the solution, but as the equations of motion conserve the phase space volume according to the Liouville theorem, the kinetic equation can be expressed in a form similar to the Kolmogorov forward equation
\begin{align}
\frac{\partial f}{\partial t}(\mathbf{z},t)=-\frac{\partial}{\partial\mathbf{z}}\cdot\left[\mathbf{a}(\mathbf{z},t)f(\mathbf{z},t)\right]+\frac{\partial}{\partial\mathbf{z}}\frac{\partial}{\partial\mathbf{z}}:\left[\mathbf{c}(\mathbf{z},t)f(\mathbf{z},t)\right],
\end{align}
where it is important to keep in mind that the quantity $\mathbf{a}$ contains \emph{also the equations of motion}, $\mathbf{\dot{z}}$. The Kolmogorov, or more familiarly, the Fokker-Planck equation describes how the probability density for finding a test particle at phase space location $\mathbf{z}$ evolves in time when the motion of an individual particle is determined by a stochastic differential equation
\begin{align}
dz^{\alpha}=a^{\alpha}dt+\sigma^{\alpha\beta}d\mathcal{W}^{\beta},
\end{align}
where the matrix $\sigma^{\alpha\beta}$ satisfies
\begin{align}
c^{\alpha\beta}=\frac{1}{2}\sigma^{\alpha\gamma}\sigma^{\beta\gamma},
\end{align}
and $\mathcal{W}^{\alpha}$ are independent stochastic Wiener processes with zero mean. This connection between stochastic processes and partial differential equations has been known since the work of Kolmogorov~\cite{math_ann:104::415:1931,kolmogorov1933grundbegriffe}. The solution to the kinetic equation is then obtained simulating random paths and taking a statistical average of them. 

In a strong magnetic field, the charged particle orbit consists of rapid gyro motion around the magnetic field lines, combined with various drifts brought about by the non-uniformity of the field or by the combined effect of magnetic and electric fields. The basic equations of motion contain all this physics. In fusion-related applications, it is often necessary to follow particles for millions of oscillation periods and the computational effort of numerical integration then becomes rather expensive. Furthermore, in cases where the background quantities are practically constant across the Larmor radius (i.e., have large gradient lengths), the full gyro motion is redundant. Then, a more attractive approach is the guiding center transformation of the charged particle Lagrangian, leading to equations of motion where the rapid oscillatory gyro motion is isolated into only one variable, namely the gyro-angle, that is not needed for following the particle's guiding center. This formalism, although only an approximation, is a powerful tool and significantly reduces the computational demands for the orbit-following, when applicable.

The equations of motion in the particle phase space, corresponding to the full gyro orbit, are introduced in Sec.~\ref{sec:FO}, and the ones corresponding to the guiding center formalism are described in Sec.~\ref{sec:GC}. The algorithms for solving these equations are also explained. In Sec.~\ref{sec:FO_coll} we describe the collision operator for the particle phase space and, in Sec.~\ref{sec:GC_coll}, the corresponding operator is introduced for the guiding center phase space.

\subsection{Hamiltonian orbit-following in electromagnetic fields: Full gyro motion}
\label{sec:FO}
The Lagrangian for a particle with mass $m$ and charge $e$ under the influence of a magnetic vector potential $\mathbf{A}(\mathbf{r},t)$ and electric potential $\Phi(\mathbf{r},t)$ is~\cite{RevModPhys.81.693}
\begin{align}
\label{eq:full_lagrangian}
L=\frac{1}{2}m\mathbf{\dot{r}}\cdot\mathbf{\dot{r}}-e\Phi+e\dot{\mathbf{r}}\cdot\mathbf{A},
\end{align}
which is equivalent to the non-canonical Hamiltonian equations of motion
\begin{align}
\label{eq:full_v}
\dot{\mathbf{v}}=&\frac{e}{m}\left(\mathbf{E}+\mathbf{v}\times\mathbf{B}\right),\\
\label{eq:full_r}
\dot{\mathbf{r}}=&\mathbf{v},
\end{align}
where $\mathbf{v}$ is the particle velocity, and the electric and magnetic fields are defined by
\begin{align}
\mathbf{E}&=-\frac{\partial\mathbf{A}}{\partial t}-\nabla\Phi,\\
\mathbf{B}&=\nabla\times\mathbf{A}.
\end{align}
The change in the total energy of the particle is given by
\begin{align}
\dot{H}=e\frac{\partial\Phi}{\partial t}-e\frac{\partial\mathbf{A}}{\partial t}\cdot\mathbf{\dot{r}},
\end{align}
which is zero for time-independent potentials, but as the equations~(\ref{eq:full_v}) and (\ref{eq:full_r}) lead to a very rapidly oscillating motion with the characteristic frequency of $\Omega=eB/m$, implementing a numerical integration scheme that would have the same property over long time periods poses a challenge.

Advancing the equations of motion in time is the essence of the ASCOT code. For the full gyro motion the code has two different options: a fourth-order Runge-Kutta method with fifth-order error estimation~\cite{Press1986}, and a modified leap-frog method. With rapidly oscillating non-canonical systems, e.g., Eqs.~(\ref{eq:full_v}) and~(\ref{eq:full_r}), the Runge-Kutta method, however, has the well-known tendency to cause numerical drift in the total energy. In contrast, a modified leap-frog method defined by
\begin{align}
\begin{split}
\label{eq:leap-frog}
\mathbf{v}_{i+1}=&\mathbf{v}_i+\Delta t\frac{e}{m}\left(\mathbf{E}_{i}+\frac{\mathbf{v}_{i+1}+\mathbf{v}_{i}}{2}\times\mathbf{B}_{i}\right),\\
\mathbf{r}_{i+1}=&\mathbf{r}_{i}+\Delta t\mathbf{v}_{i},
\end{split}
\end{align}
has the property that $\mathbf{v}_{i+1}\cdot\mathbf{v}_{i+1}=\mathbf{v}_{i}\cdot\mathbf{v}_{i}$ if the electric field is zero, i.e., it explicitly conserves energy.

\subsection{Hamiltonian orbit-following in electromagnetic fields: Guiding center motion}
\label{sec:GC}
For a guiding center phase space $Z^{\gamma}=(\mathbf{R},\mathrm{v}_{\parallel},\mu,\chi)$, i.e, location, parallel velocity, magnetic moment, and gyro-angle, respectively, the transformed Lagrangian is~\cite{RevModPhys.81.693}
\begin{align}
\label{eq:gc_lagrangian}
L=(e\mathbf{A}+m\mathrm{v}_{\parallel}\mathbf{b})\cdot\dot{\mathbf{R}}+\frac{m\mu}{e}\dot{\chi}-H,
\end{align}
where $\mathbf{b}$ is the magnetic field unit vector, and the Hamiltonian $H$ is given by
\begin{align}
\label{eq:gc_hamiltonian}
H=\frac{1}{2}m\mathrm{v}_{\parallel}^2+\mu B+e\Phi=\mathcal{E}+e\Phi.
\end{align}
The guiding center equations of motion are obtained applying the Euler-Lagrange equation $\frac{d}{dt}\left(\frac{\partial L}{\partial\dot{Z}^{\gamma}}\right)-\frac{\partial L}{\partial Z^{\gamma}}=0$ for each phase space coordinate, yielding
\begin{align}
\label{eq:chidot}
\dot{\chi}=&\frac{eB}{m},\\
\label{eq:mudot}
\dot{\mu}=&0,\\
\label{eq:vpardot}
\dot{\mathrm{v}}_{\parallel}=&\frac{e}{m}\frac{\mathbf{B}^{\star}}{B_{\parallel}^{\star}}\cdot\mathbf{E}^{\star},\\
\label{eq:rdot}
\dot{\mathbf{R}}=&\mathrm{v}_{\parallel}\frac{\mathbf{B}^{\star}}{B_{\parallel}^{\star}}+\mathbf{E}^{\star}\times\frac{\mathbf{b}}{B_{\parallel}^{\star}},
\end{align}
where $B_{\parallel}^{\star}=\mathbf{B}^{\star}\cdot\mathbf{b}$, and the effective fields ($\mathbf{E}^{\star}=-\partial\mathbf{A}^{\star}/\partial t-\nabla\Phi^{\star}$, $\mathbf{B}^{\star}=\nabla\times\mathbf{A}^{\star}$) in Eqs.~(\ref{eq:vpardot}) and~(\ref{eq:rdot}) are defined by the effective potentials
\begin{align}
\label{eq:Phistar}
\Phi^{\star}(\mathbf{R},\mu,t)=\Phi+\mu B/e,\\
\label{eq:Astar}
\mathbf{A}^{\star}(\mathbf{R},\mathrm{v}_{\parallel},t)=\mathbf{A}+m\mathrm{v}_{\parallel}\mathbf{b}/e.
\end{align}
The time rate of change in the total energy for the guiding center then becomes
\begin{align}
\begin{split}
\label{eq:hdot}
\dot{H}=e\frac{\partial\Phi^{\star}}{\partial t}-e\frac{\partial\mathbf{A}^{\star}}{\partial t}\cdot\mathbf{\dot{R}},
\end{split}
\end{align}
and, for a static background, it is zero, as expected.

The oscillation of the guiding center orbit happens in a radically different time scale than that of the gyro motion, and a fourth-order Runge-Kutta method with fifth-order error checking~\cite{Press1986} has proved itself adequate when numerically integrating the guiding center equations of motion.

\subsection{Coulomb collisions in particle phase space}
\label{sec:FO_coll}
The Coulomb collisions are modelled by operators acting in particle velocity space. The collisional part of the kinetic equation takes the form
\begin{align}
\left(\frac{\partial f}{\partial t}\right)_{coll}=\;-\frac{\partial}{\partial\mathbf{v}}\cdot\left[\mathbf{A}f-\frac{1}{2}\frac{\partial}{\partial\mathbf{v}}\cdot\left(\mathbf{D}f\right)\right],
\label{eq:perturbations_collision_term}
\end{align}
where the vector $\mathbf{A}$ can be interpreted as a friction term in velocity space and the tensor $\mathbf{D}$ as a velocity diffusion term. Their explicit expressions are
\begin{align}
\label{eq:perturbations_friction_term}
\mathbf{A}&=-2\sum_bc_{b}\left(1+\frac{m}{m_b}\right)\int d\mathbf{v}'f_b(\mathbf{v}')\frac{\mathbf{u}}{u^3}\\
\label{eq:perturbations_diffusion_term}
\mathbf{D}&=2\sum_bc_{b}\int d\mathbf{v}'f_b(\mathbf{v}')\left(\frac{\mathbf{I}}{u}-\frac{\mathbf{u}\mathbf{u}}{u^3}\right),
\end{align}
where $c_{b}=e^2e^2_b\ln\Lambda/(8\pi\epsilon_0^2m^2)$, $\ln\Lambda$ is the Coulomb logarithm, $e_b$ is the electric charge of the plasma species $b$, and $\mathbf{u}=\mathbf{v}-\mathbf{v}'$. For a comprehensive derivation of the coefficients, see~\cite{ichimaru1973basic}.

In the case of a Maxwellian background plasma with no flow velocity, the friction and diffusion coefficients are reduced to the form
\begin{align}
\mathbf{A}&=\nu_s\mathbf{v},\\
\mathbf{D}&=D_{\parallel}\frac{\mathbf{v}\mathbf{v}}{v^2}+D_{\perp}\left(\mathbf{I}-\frac{\mathbf{v}\mathbf{v}}{v^2}\right),
\end{align}
where the scalar coefficients are
\begin{align}
D_{\parallel}(\mathrm{v})&=\sum_b 4n_bc_{b}\sqrt{\frac{m_b}{2kT_b}}\frac{\Psi(x)}{x},\\
D_{\perp}(\mathrm{v})&=\sum_b2n_bc_{b}\sqrt{\frac{m_b}{2kT_b}}\left(\Phi(x)-\Psi(x)\right),\\
\nu_s(\mathrm{v})&=\sum_b\frac{4n_bc_{b}m_b}{kT_b}\left(1+\frac{m}{m_b}\right)\Psi(x),
\end{align}
and $x=\mathrm{v}/\sqrt{2kT_b/m_b}$. The functions $\Phi$ and $\Psi$ are defined by
\begin{align}
\Phi(x)&=\frac{2}{\sqrt{\pi}}\int_0^x \exp{(-y^2)}dy,\\
\Psi(x)&=\frac{\Phi(x)-x\Phi'(x)}{2x^2}.
\end{align}

The particle motion is then governed both by the Hamiltonian motion, Eq.~(\ref{eq:full_v}), and by Coulomb drag and diffusion, yielding for the velocity
\begin{align}
\label{eq:dpprt}
d\mathbf{v}\;=&\;\left[\frac{e}{m}\left(\mathbf{E}+\mathbf{v}\times\mathbf{B}\right)+\nu_s\mathbf{v}\right]dt\nonumber\\&+\left[\sqrt{D_{\parallel}}\frac{\mathbf{v}\mathbf{v}}{v^2}+\sqrt{D_{\perp}}\left(\mathbf{I}-\frac{\mathbf{v}\mathbf{v}}{v^2}\right)\right]\cdot d\mathbf{\mathcal{W}}^{\mathbf{v}},
\end{align}
while the particle location obeys Eq.~(\ref{eq:full_r}). The deterministic part of Eq.~(\ref{eq:dpprt}), containing both the Hamiltonian equation of motion and the collisional friction term, can be advanced in time using the leap-frog scheme introduced in Sec.~\ref{sec:FO}, while the stochastic part is often treated with the stochastic Euler method.

\subsection{Coulomb collisions in the guiding center phase space}
\label{sec:GC_coll}
It is essential to understand that the guiding center transformation should be applied consistently to the whole kinetic equation,  not just to the equations of motion. An exclusive guiding center transformation of the kinetic equation with Coulomb collisions included is given in~\cite{brizard:fokkerPlanck}, and the corresponding stochastic differential equations for guiding center phase space coordinates are discussed in~\cite{hirvijoki:092505:2013:pop}. Although the formalism includes the effects of magnetic drifts, within this work, the guiding center Coulomb collision are considered only up to zeroth order in magnetic field non-uniformity. The most evident consequence of the proper treatment of the guiding center Coulomb collisions, the appearance of a {\emph spatial} diffusion in addition to the velocity space operators, however, is present also in zeroth order.

The simplest representation of the operators involves the guiding center velocity $\mathrm{v}=\sqrt{2\mathcal{E}/m}$ and pitch $\xi=\mathrm{v}_{\parallel}/\mathrm{v}$, for which the zeroth order Coulomb contributions to the stochastic differential equations are
\begin{align}
\label{eq:gcdv}
d\mathrm{v}_{coll}&=\left(-\nu\mathrm{v}+\frac{\partial D_{\parallel}}{\partial\mathrm{v}}+2\frac{D_{\parallel}}{\mathrm{v}}\right)dt+\sqrt{2D_{\parallel}}d\mathcal{W}^v\\
\label{eq:gcdxi}
d\xi_{coll}&=-\xi\frac{2D_{\perp}}{v^2}dt+\sqrt{\left(1-\xi^2\right)\frac{2D_{\perp}}{\mathrm{v}^2}}d\mathcal{W}^{\xi}.
\end{align}
Here the frequency $\nu$ differs slightly from $\nu_s$, and is defined by
\begin{align}
\nu(\mathrm{v})=\sum_b\frac{4n_bc_{b}m_b}{kT_b}\Psi(x).
\end{align}

The guiding center equations of motion, however, are often given for the velocity space coordinates $(\mathrm{v}_{\parallel},\mu)$, and solving the kinetic equation by the stochastic approach calls for a consistent phase space. For the coordinates $(\mathrm{v}_{\parallel},\mu)$ the zeroth order stochastic equations become
\begin{align}
\label{eq:vpar_stoc}
d\mathrm{v}_{\parallel}\;=\;&\left[\dot{\mathrm{v}}_{\parallel}-\nu\mathrm{v}_{\parallel}+\xi\left(2\frac{D_{\parallel}-D_{\perp}}{\mathrm{v}}+\frac{\partial D_{\parallel}}{\partial\mathrm{v}}\right)\right]dt\nonumber\\&+\Sigma^{\mathrm{v}_{\parallel}\mathrm{v}_{\parallel}}d\mathcal{W}^{\mathrm{v}_{\parallel}}+\Sigma^{\mathrm{v}_{\parallel}\mu}d\mathcal{W}^{\mu}\\
\label{eq:mu_stoc}
d\mu\;=\;&\left[-2\nu\mu+\frac{m\mu}{\mathcal{E}}\left(\mathrm{v}\frac{\partial D_{\parallel}}{\partial\mathrm{v}}+3\left(D_{\parallel}-D_{\perp}\right)\right)+\frac{2mD_{\perp}}{B}\right]dt\nonumber\\&+\Sigma^{\mu\mathrm{v}_{\parallel}}d\mathcal{W}^{\mathrm{v}_{\parallel}}+\Sigma^{\mu\mu}d\mathcal{W}^{\mu},
\end{align}
where, in Eq.~(\ref{eq:vpar_stoc}), $\dot{\mathrm{v}}_{\parallel}$ is given by Eq.~(\ref{eq:vpardot}), and we have neglected $\dot{\mu}$ since the magnetic moment is a constant of Hamiltonian guiding center motion. The matrix $\Sigma^{\alpha\beta}$ is solved from the condition $\Sigma^{\alpha\gamma}\Sigma^{\beta\gamma}=2\mathcal{D}^{\alpha\beta}$, where the symmetric guiding center velocity space diffusion matrix $\mathbf{\mathcal{D}}=\mathbf{\mathcal{B}}\mathbf{\mathcal{Y}}\mathbf{\mathcal{B}}$ is expressed as a product where the diagonal matrix $\mathbf{\mathcal{B}}$ has the components
\begin{align}
\mathcal{B}^{v_{\parallel}v_{\parallel}}=&1,\\
\mathcal{B}^{\mu\mu}=&\mathcal{E}/(\mathrm{v}B),
\end{align}
and the symmetric normalized matrix $\mathbf{\mathcal{Y}}$ has the components
\begin{align}
\mathcal{Y}^{v_{\parallel}v_{\parallel}}&=\left[ D_{\parallel}\xi^2+D_{\perp}(1-\xi^2)\right]\\
\mathcal{Y}^{v_{\parallel}\mu}&=2\xi (1-\xi^2)(D_{\parallel}-D_{\perp}) \\
\mathcal{Y}^{\mu\mu}&=4(1-\xi^2)\left[D_{\parallel}(1-\xi^2)+D_{\perp}\xi^2\right].
\end{align}

The zeroth order Coulomb contribution to the spatial guiding center coordinates appears as a purely diffusive process, yielding for the guiding center position
\begin{align}
\label{eq:spatial}
d\mathbf{R}=\mathbf{\dot{R}}dt+\sqrt{2D_c^{\mathbf{R}}}\left(\mathbf{I}-\mathbf{b}\mathbf{b}\right)\cdot d\mathbf{\mathcal{W}}^{\mathbf{R}},
\end{align}
where the spatial Coulomb diffusion coefficient is
\begin{align}
D_c^{\mathbf{R}}=\frac{1}{2\Omega^2}\left[D_{\parallel}(1-\xi^2)+D_{\perp}(1+\xi^2)\right],
\end{align}
and $\mathbf{\dot{R}}$ is given by Eq.~(\ref{eq:rdot}).

As in the case of gyro orbit following, the deterministic part of the collisional effects should be evaluated together with the equations of motion, using the higher level integrators, while the stochastic part can be evaluated using the Euler method.

\section{Fast ion sources}
\label{sec:sources}
The ASCOT code is often used to gather information from fast ions. For convenience, the code has an in-built capability to initialize test particles that represent the actual particles. The sources listed below include energetic ions and neutrons from fusion reactions, as well as ions generated by neutral beam injection or ICRH acceleration, all relevant for studies of, e.g., plasma heating and current drive or fast ion power loads to plasma facing components. 

\subsection{Fusion product source}
\label{sec:fus}

Generating test particles that represent the fusion products is a two-phase process. First, the reaction rates for the fusion reactions of interest are calculated onto a cylindrical grid $\mathcal{R}(R,z)$. Then, test particles are generated making use of this grid. From here on, the symbol {z} no longer stands for the phase space coordinate but, rather, has its conventional role  as one of the cylindrical coordinates.

The four reactions described by Bosch and Hale\,\cite{0029-5515-32-4-I07} are supported: D(d,n)$^3$He, D(d,p)T, T(d,n)$\alpha$ and $^3$He(d,P)$\alpha$. These reactions are considered in three different cases: thermal, beam-target and beam-beam. For the thermal case Bosch and Hale provide a parametrized model. The background plasma parameters are used directly from the input files. The beam-target and beam-beam reaction rates are calculated as a post-processing step where the fast ion density distribution $n_i(R,z,\mathrm{v}_\parallel,\mathrm{v}_\perp)$ from an earlier run reacts either with itself (beam-beam) or with the background plasma (beam-target).

These calculations are integrals over one or two fast ion distributions. To expedite the integral, intermediate averaged reactivities are calculated. For the beam-beam reaction the fusion cross section is first averaged over the gyro angle $\chi$ in a similar way that NUBEAM~\cite{Goldston198161,Pankin2004157} does in order to get the averaged reactivity
\begin{equation}
  \label{eq:bb-average}
  \left < \sigma\mathrm{v}\right >_\textrm{BB}(\mathrm{v}_{i\perp},\mathrm{v}_{j\perp},|\mathrm{v}_{i||}-\mathrm{v}_{j||}|)=\int \sigma(|\mathbf{g}|)|\mathbf{g}|d\chi_{i}d\chi_{j}
\end{equation}
where $\mathbf{g}=\mathbf{v}_i-\mathbf{v}_j$. Then, the beam-beam fusion rate is calculated according to
\begin{equation}
  \label{eq:beam-beam-rate}
  \mathcal{R}_\textrm{BB}=\int\left < \sigma\mathrm{v}\right >_\textrm{BB} n_i(\mathrm{v}_{i\perp},\mathrm{v}_{i||})n_j(\mathrm{v}_{j\perp},\mathrm{v}_{j||})d\mathrm{v}_{i||}d\mathrm{v}_{j||}d\mathrm{v}_{i\perp}d\mathrm{v}_{j\perp}
\end{equation}
 For the beam-target reaction, the reaction rate is given by
\begin{equation}
  \label{eq:beam-target-rate}
  \mathcal{R}_\textrm{BT}=\int n_\textrm{B}(\mathbf{v}_\textrm{B})n_\textrm{T}(\mathbf{u}+\mathbf{v}_\textrm{B})\sigma(|\mathbf{u}|)|\mathbf{u}|d\mathbf{u}d\mathbf{v}_\textrm{B},
\end{equation}
where $n_\textrm{B}$ is the fast ion density calculated by ASCOT and $n_\textrm{T}$ is the Maxwellian density distribution of the background plasma.

Once the reaction rate $\mathcal{R}$ (reactions per volume and time) has been calculated, there are two methods to initialize fusion products: a weighted initialization and a uniform weight initialization. In the weighted initialization, the coordinates ($R,\phi,z$) are chosen by uniformly distributed random numbers between the minimum and maximum dimensions set by the walls and the X-point. If the chosen location is not inside the separatrix, the particle is discarded and a new location is picked. Once a location inside the separatrix is found, a weight factor is calculated as
\begin{equation}
w_i = \int_0^{2\pi}\mathcal{R}(\rho_{p,i})d\phi \frac{\int_0^1\mathcal{R}(\rho_p) V(\rho) d\rho_p}{\sum^N_{i=1} \int_0^{2\pi}\mathcal{R}(\rho_{p,i}) d\phi},
\label{eq:weight}
\end{equation}
where $w_i$ is the weight of the $i$th particle, $\mathcal{R}(\rho_p)$ is the fusion reaction rate with $\mathcal{R}(\rho_{p,i})$ being the rate evaluated at the location of the {\it i}th particle, $N$ is the total number of particles, $V(\rho_p)$ is the flux surface volume, $\rho_p = \sqrt{\psi_p}$ is the square root of the normalized poloidal magnetic flux, and $\phi$ is the toroidal angle. The initial velocity is chosen so that the velocity distribution is isotropic.

In the uniform weight initialization, R, $\phi$ and z are chosen as random numbers uniformly distributed in volume, between the minimum and maximum dimensions set by the walls and the X-point. Then a uniformly distributed random number $X$ between $0$ and $1$ is compared against the normalized local reaction rate $\mathcal{R}_N = \mathcal{R}/\max(\mathcal{R})$. If $X \leq \mathcal{R}_N$, the particle is generated. The weight factors for all particles are now equal, calculated simply as
\begin{equation}
w_i = w = \frac{\int_0^1\mathcal{R}(\rho_p) V(\rho) d\rho_p}{N}.
\end{equation}

\subsection{Neutral beam ion source}
\label{sec:nbi}
Due to the structure of neutral beam injectors used in fusion devices worldwide, the produced beams consist of small sub-beams, or beamlets. In order to model neutral beam injection (NBI) as accurately as possible, this fine structure of the beam is taken into account in the NBI ion source code used in ASCOT.

To generate an NBI test particle, a neutral particle from a random beamlet is chosen. The neutral is assigned a velocity in the direction of the beamlet, offset by a usually bi-gaussian dispersion, and advanced along its velocity vector until it either hits an obstacle or enters the vacuum chamber. Once inside the vessel, the neutral particle is given a uniformly distributed random threshold value $\lambda~\in~[0, 1]$ and advanced along a straight trajectory until the probability, calculated for the particle to survive further without ionization, is lower than the threshold~$\lambda$.

A general probabilistic model for effective ionization and recombination is derived considering a large number of particles, $N$, with some initial state, and assuming that each reaction $s$ removes particles from the initial state according to the differential equations
\begin{align}
\frac{dN_s}{dx}&=\Sigma_s(x)N(x),\\
\frac{dN}{dx}&=-\sum_s\frac{dN_s}{dx}
\end{align}
where $N_s$ is the number of reactions of type $s$ that have taken place before reaching the location $x$ and $\Sigma_s(x)$ is the cross-section. Integration gives the solution
\begin{align}
N_s(x)&=N_0\int_0^x\Sigma_s(x')\exp{\left(-\int_0^{x'}\sum_k\Sigma_k(x'')dx''\right)}dx'\\
\label{eq:decay}
N(x)&=N_0\exp{\left(-\int_0^x\sum_k\Sigma_k(x')dx'\right)},
\end{align}
which satisfies $\sum_sN_s(x)+N(x)=N_0$ and, thus, conserves the number of particles. Letting $\mathcal{X}_s$ denote a random location for the reaction $s$ to happen, the probability for ''reaction $s$ takes place before reaching location $x$'' is
\begin{align}
\label{eq:prop_model}
Pr(\mathcal{X}_s\leq x)=\int_{0}^{x}R_s(x')\exp{\left(-\int_0^{x'}\sum_k\Sigma_k(x'')dx''\right)}dx'.
\end{align}

The model for a neutral to become ionized involves only the effective ionization, and the cumulative probability for a neutral to survive the interval $[0,x_i]$ is thus
\begin{align}
\label{eq:Pneutr}
P_{i} =&\exp{\left(-\int_{0}^{x_{i-1}}\Sigma(x')dx'\right)}\exp{\left(-\int_{x_{i-1}}^{x_i}\Sigma(x')dx'\right)}\nonumber\\
=&P_{i-1} \exp{\left(-\int_{x_{i-1}}^{x_i}\Sigma(x')dx'\right)},
\end{align}
where $\Sigma$ is the total ionization cross-section calculated from the analytical fits given by Suzuki {\it et al.}~\cite{0741-3335-40-12-009}, and $P_{i-1}$ is the probability to survive the interval $[0,x_{i-1}]$. In the code the integral is discretized with small steps $\Delta x_i=x_i-x_{i-1}$, yielding
\begin{align}
P_{i} = P_{i-1} \exp{\left(-\Sigma_{i}\Delta x_i\right)},
\end{align}
and once $P_{i}\leq\lambda$ is obtained, i.e., the probability of the particle surviving to its current location falls below the threshold, the particle is backtracked by a fraction of the last step to the exact location where the exponential crosses the threshold value, given by
\begin{align}
\Delta x_{i}=-\frac{1}{\Sigma_i}\ln{\frac{\lambda}{P_{i-1}}}.
\end{align}
At this location, a test particle is recorded. If the neutral particle hits the wall of the device before being ionized, it is considered shine-through.

At the time of writing, the NBI geometries of JET, ITER, ASDEX Upgrade, DIII-D, FAST, TEXTOR, MAST, and Tore Supra have already been implemented. Adding the NBI geometries of new devices and benchmarking the model against existing codes is an ongoing project. A more detailed description of the NBI model and comparisons to other codes will be presented in a separate publication.

\subsection{Ion cyclotron resonance heated ion source}
\label{sec:icrh}

Ion Cyclotron Resonance Heated (ICRH) ions can be considered at different levels of sophistication. The most sophisticated level requires a self-consistent simulation taking into account the wave field caused by the ICRH antenna and its interaction with the plasma. Currently only a few dedicated models \cite{Hedin:Nucl:Fusion:2002:selfo,Hellsten:Nucl:Fusion:2004:selfo,Brambilla:1999:0741-3335:1:toric_code,Wright:IEEE:38:2010:torlh_code,JuckerMartinPhdScenic} exist that can provide this kind of a realistic ICRH distribution. Often, however, it is enough to use approximations, e.g., when one is looking for the types of particle orbits that will be lost to plasma facing components.

The ICRH ion source model for ASCOT is based on physical observations. The plasma acts as a lens for the wave field, focusing it on the magnetic axis. Due to the effects of, e.g., finite wavelength, finite absorptivity and up-down asymmetry, the focusing is not perfect, and the distribution of ICRH ions is peaked at the magnetic axis with a finite half-width responsible for spreading in the radial coordinate $\rho_{p}$. In addition, the ICRH ions will have the banana turning point at places where the frequency of the ICRH wave, $\omega$, meets the resonance condition, $\omega = n\Omega$, for the $n$th harmonic of the wave field. As $\Omega$ is proportional to the magnetic field $B$, which is roughly a function of the inverse major radius $1/R$, the ICRH distribution will be roughly limited to a certain resonant major radius rather than spreading in $\rho_p$ only. 

The implementation in ASCOT first samples a Gaussian distribution, with a user-defined peak location, $\rho_{\textrm{peak}}$, and half-width, $\rho_{\textrm{hw}}$, to get an ensemble of $\rho$ values. The energies are sampled from a Maxwellian distribution with a given temperature, e.g. 500 keV, and an optional cut-off energy $E_{c\mbox{-}o}$. After this, the $(R,z)$ locations for the particles are found by minimizing the function
\begin{equation}
f(R,z) = a\left[\rho_p(R,z)-\rho\right]^2+b\left[n\Omega(R,z)-\omega\right]^2,
\end{equation}
where $a$ and $b$ are user-defined constant parameters. Finally, the particles are given a weight factor, scaled according to the total ICRH power $P_{\rm ICRH}$ and the number of test particles created.

\section{Magnetohydrodynamic activity for fast ion modelling}
\label{sec:mhd}
High performance fusion plasmas rarely are MHD quiescent. In particular, all ITER plasmas are prone to neoclassical tearing modes (NTM), and the advanced scenario plasmas corresponding to steady-state operation are likely to exhibit significant Alfv\'{e}n activity. Both of these MHD phenomena are expected to affect the fast ion trajectories and, therefore, possibly lead to redistribution of the fast ion population. To account for the physics related to such MHD activity, a new model, applicable in 3D geometries, is incorporated in ASCOT.

The model for both stationary neoclassical tearing modes and rotating Alfv\'{e}n eigenmodes (AE) is reported in detail in~\cite{mhdpapru}. The MHD activity is introduced as a perturbation in the magnetic vector potential, $\tilde{\mathbf{A}}=\alpha\mathbf{B}$, while a possible time dependence generates a perturbation also in the electric potential, $\tilde{\Phi}$. As the modes often appear as harmonic structures along the magnetic field lines, a helical structure is assumed for the perturbations
\begin{align}
\alpha&=\sum_{nm}\alpha_{nm}(\psi_p)\sin{\left(n\zeta-m\theta-\omega_{nm} t \right)} \\
\tilde{\Phi}&=\sum_{nm}\Phi_{nm}(\psi_p)\sin{\left(n\zeta-m\theta-\omega_{nm} t \right)},
\end{align}
and Boozer coordinates ($\psi_p,\theta,\zeta$) are currently used to map the perturbative quantities back to cylindrical coordinates.

For NTMs, the rotation frequency $\omega_{nm}$ is usually low, and we may assume no electric perturbation to appear, thus neglecting $\tilde{\Phi}$ and $\omega_{nm}$. In the case of a rapidly rotating AE, the approximation that the perturbed parallel electric field vanishes due to the rapid motion of electrons along the field line implies that $\alpha_{nm}(\psi_p)$ and $\Phi_{nm}(\psi_p)$ are related, and only one of them is required to describe the mode in the Boozer coordinates~\cite{white2006theorytoroidalplasmas}. The input data for ASCOT can be obtained, e.g., from MHD codes like LIGKA~\cite{LIGKA}, and the data typically consists of radial functions $\Phi_{nm}$, mode numbers $n$ and $m$, and frequencies $\omega_{nm}$.

Although the model is strongly dependent on the coordinates ($\psi_p,\theta,\zeta$), it differs from previous work: the test particle guiding center can be followed in any coordinate system, contrary to previous methods restricted to field-aligned coordinates only. Adding the electric perturbation causes no changes to the equations of motion: only the gradient of the perturbation potential is added into the electric field that already appears in the equations. Proceeding similarly with the magnetic vector potential would require calculating terms like $\nabla\times\nabla\times (\alpha\mathbf{B})$. This is rendered unnecessary by adding the magnetic perturbation only into the symplectic part of the Lagrangian, Eq.~(\ref{eq:gc_lagrangian}), and neglecting it from the Hamiltonian. As a result, we introduce the modified potentials ($\mathbf{A}^{\star\star},\Phi^{\star\star}$) which are related to the effective potentials ($\mathbf{A}^{\star},\Phi^{\star}$) by
\begin{align}
\mathbf{A}^{\star\star}&=\mathbf{A}^{\star}+\alpha\mathbf{B},\\
\Phi^{\star\star}&=\Phi^{\star}+\tilde{\Phi}.
\end{align}
Thus, in Eqs.~(\ref{eq:vpardot}) and~(\ref{eq:rdot}), the effective fields $\mathbf{B}^{\star}$ and $\mathbf{E}^{\star}$ are replaced with
\begin{align}
\mathbf{B}^{\star\star}&=\nabla\times\mathbf{A}^{\star\star},\\
\mathbf{E}^{\star\star}&=-\frac{\partial\mathbf{A}^{\star\star}}{\partial t}-\nabla\Phi^{\star\star}.
\end{align}
and also the time rate of change of the total energy is now expressed by the modified potentials
\begin{align}
\begin{split}
\label{eq:hdotmhd}
\dot{H}=e\frac{\partial\Phi^{\star\star}}{\partial t}-e\frac{\partial\mathbf{A}^{\star\star}}{\partial t}\cdot\mathbf{\dot{R}}
\end{split}
\end{align}
still yielding zero for static backgrounds and time-independent perturbations.

The implementation of the time-dependent method thus requires $\alpha$, $\nabla\alpha$, $\nabla\tilde{\Phi}$ and $\partial\alpha/\partial t$ at given location. These quantities are calculated with the aid of coordinate transformations from cylindrical/Cartesian to Boozer coordinates and back.

\section{Interaction models for impurity particles}
\label{sec:impurity}
In addition to fast particles, ASCOT can also be used for modelling impurity transport. The code has been applied to, e.g., simulating trace element injection experiments~\cite{miettunen:NF2012}. For these purposes, the advantage of ASCOT is impurity following in all regions of the plasma, extending from the core plasma, scrape-off layer (SOL) and halo plasma to the wall in a fully 3D tokamak environment.

Compared to fast particles, however, impurities in the SOL have typically very low energies (of the order of 1--100 eV). Due to the low energy, impurities are more strongly affected by the background plasma properties, such as the flow velocity. Additionally, the charge state of impurity particles can vary significantly as a result of their high $Z$ number, which further affects their transport. Therefore, additional interaction models are needed for realistic simulations of impurities.

\subsection{Background plasma flow}
Strong plasma flows with typical velocities of Mach 0.5--1 have been measured in the SOL region of various tokamaks~\cite{Asakura200741:SOLflow}. Owing to the low energy of impurities, the flow has a significant effect on their long range transport. The Coulomb collision operators introduced in Sec.~\ref{sec:equations}, however, assume the background plasma to have a purely Maxwellian distribution, which neglects plasma flow.

In ASCOT simulations, the effect of background plasma flow on Coulomb collisions can be modelled by using a frame of reference moving with the local parallel flow velocity $\mathrm{v}_{\parallel,\text{flow}}$ of the background plasma. As the collisions are about to be evaluated, the frame of reference is switched by updating the particle velocity according to $\mathbf{v}'=\mathbf{v}-\mathbf{\hat{b}}\mathrm{v}_{\parallel,\text{flow}}$ and calculating the Coulomb contribution in Eq.(~\ref{eq:dpprt}) with $\mathbf{v}'$ instead of $\mathbf{v}$. This procedure is equal to calculating the Coulomb coefficients with drift Maxwellian backgrounds.

For a guiding center, the parallel velocity is changed according to $\mathrm{v}_\parallel' = \mathrm{v}_\parallel-\mathrm{v}_{\parallel,\text{flow}}$, yielding the velocity $\mathrm{v}'=\sqrt{(\mathrm{v}_{\parallel}')^2+2\mu B/m}$ and the pitch $\xi'=\mathrm{v}_{\parallel}'/\mathrm{v}'$. Using either of the quantities $(\mathrm{v}',\xi')$ or $(\mathrm{v}_{\parallel}',\mu)$, the Coulomb contribution is then integrated using either Eqs.~(\ref{eq:gcdv}) and~(\ref{eq:gcdxi}) or Eqs.~(\ref{eq:vpar_stoc}) and~(\ref{eq:mu_stoc}) that give new values for $(\xi',\mathrm{v}')$ or $(\mathrm{v}_{\parallel}',\mu)$, which are then transformed back to the laboratory frame of reference, giving new values of the velocity space coordinates. 

\subsection{Atomic reactions}
During a single time step, test particles can experience several different atomic reactions, such as impact ionization, when interacting with the background plasma. These reactions change the charge state of the followed particles, which then affects their transport directly through the equations of motion described in Sections~\ref{sec:FO} and \ref{sec:GC}. For heavy impurities, such as tungsten with $Z = 74$, the effect should not be neglected.

If only the changes in the charge state of the followed particle are of interest, it is not necessary to model all possible atomic reactions individually. Instead, it is sufficient to model the effective ionization and recombination of the particle. During a single time step, the particle may undergo several reactions, which effectively lead to the ionization or recombination of the particle, or to a situation where the charge state remains effectively unchanged.

As charged particles in a magnetic field do not move linearly, it is more convenient to use time $t$ and reaction rates $R_s$ rather than the distance $x$ and cross sections $\Sigma_s$ used in the NBI model in Sec.~\ref{sec:nbi}. The probabilistic model, however, remains unchanged. Thus, letting $\mathcal{T}_s$ denote a random time for reaction $s$ to take place during time step $\Delta t$, Eq.~(\ref{eq:prop_model}) is approximated as
\begin{align}
Pr(\mathcal{T}_s\leq\Delta t)=\frac{R_s}{\sum_kR_k}\left[1-\exp{\left(-\sum_kR_k\Delta t\right) } \right]
\end{align}
and tested against a uniformly distributed random number $\lambda~\in~[0,1]$ to determine whether the charge state is increased or decreased by one or remains unchanged during the time step $\Delta t$.

The reaction rate coefficients as a function of local electron temperature and density are taken from the ADAS database \cite{adas:database}. At the time of writing, data for carbon, beryllium, tungsten and nitrogen have been imported into ASCOT.

\section{Wall collisions}
\label{sec:wall}
\begin{figure}[!h]
\begin{center}
\includegraphics[width=0.4\textwidth]{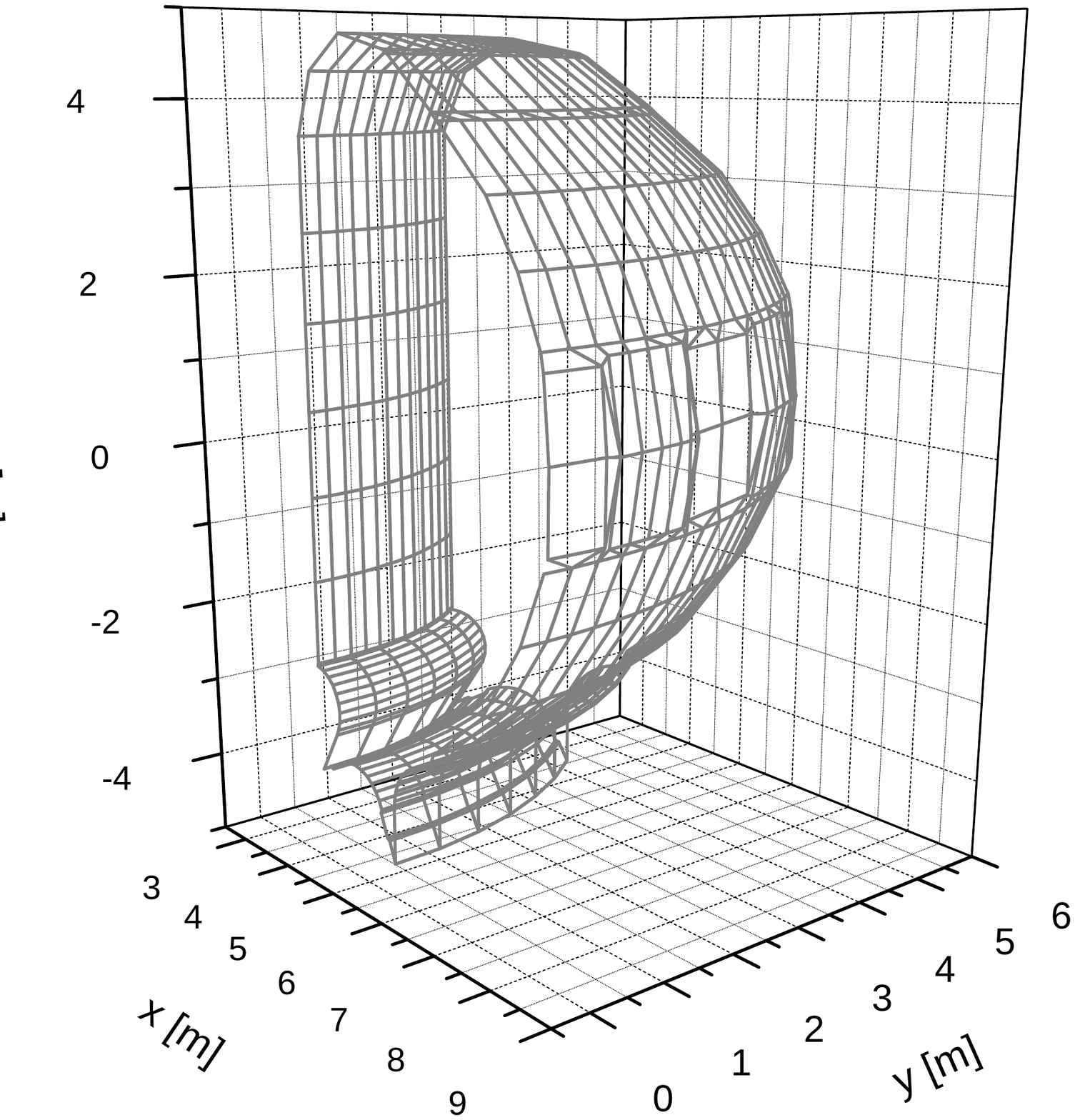}
\caption{A model ITER wall sector consisting of triangles and planar quadrangles (data provided by C. Doebert, EFDA Design Office, 2007).}
\label{fig:iter_wall_sector}
\end{center}
\end{figure}

Setting up a realistic calculation domain requires a limiting surface to be defined. If fast ion power loads or the deposition of impurities on the first wall are to be reliably simulated, this limiting surface has to accurately represent the wall structures. In ASCOT this surface is taken to be the first wall of the vessel, represented by triangular and planar quadrilateral elements as shown in Figure~\ref{fig:iter_wall_sector}. Inside the code all given planar quadrangular polygons are split into two triangles, as these are the simplest planar elements for collision detection. For each triangle, the unit normal vector $\mathbf{N}$ and an implicit presentation of the triangle's plane $\mathbf{N} \cdot \mathbf{P} + d = 0$ are defined and stored. Here, $d = -\mathbf{V} \cdot \mathbf{N}$ and $\mathbf{V}$ is any vertex of the triangle. These definitions are used by the wall collision detection algorithm during simulations.

 For modelling a 3D wall and test particle wall collisions, an efficient ray-polygon collision detection algorithm \cite{Badouel:1990:ERI:90767.90867} has been adopted in ASCOT. The test particle orbit step from $\mathbf{r}_0$ to $\mathbf{r}_1 = \mathbf{r}_0 + \Delta \mathbf{r}$, to be tested for wall collision, is presented parametrically as $\mathbf{r}(t) = \mathbf{r}_0 + t \Delta \mathbf{r}$. The parameter $t$ is evaluated at the intersection of the plane of each preliminary candidate triangle as $t = -(d + \mathbf{N} \cdot \mathbf{r}_0) / (\mathbf{N} \cdot \Delta \mathbf{r})$ and stored as $t_{i}$. All triangles for which $0 < t_{i} \le 1$ are stored as collision candidates. If no such triangles are encountered, the wall collision check routine returns a no-collision result. If collision candidate triangles are found, the plane intersection point $\mathbf{P}  = \mathbf{r}_0 + t_{i} \Delta \mathbf{r}$ is evaluated for each one of them.

\begin{figure}[!h]
\begin{center}
\includegraphics[width=0.36\textwidth]{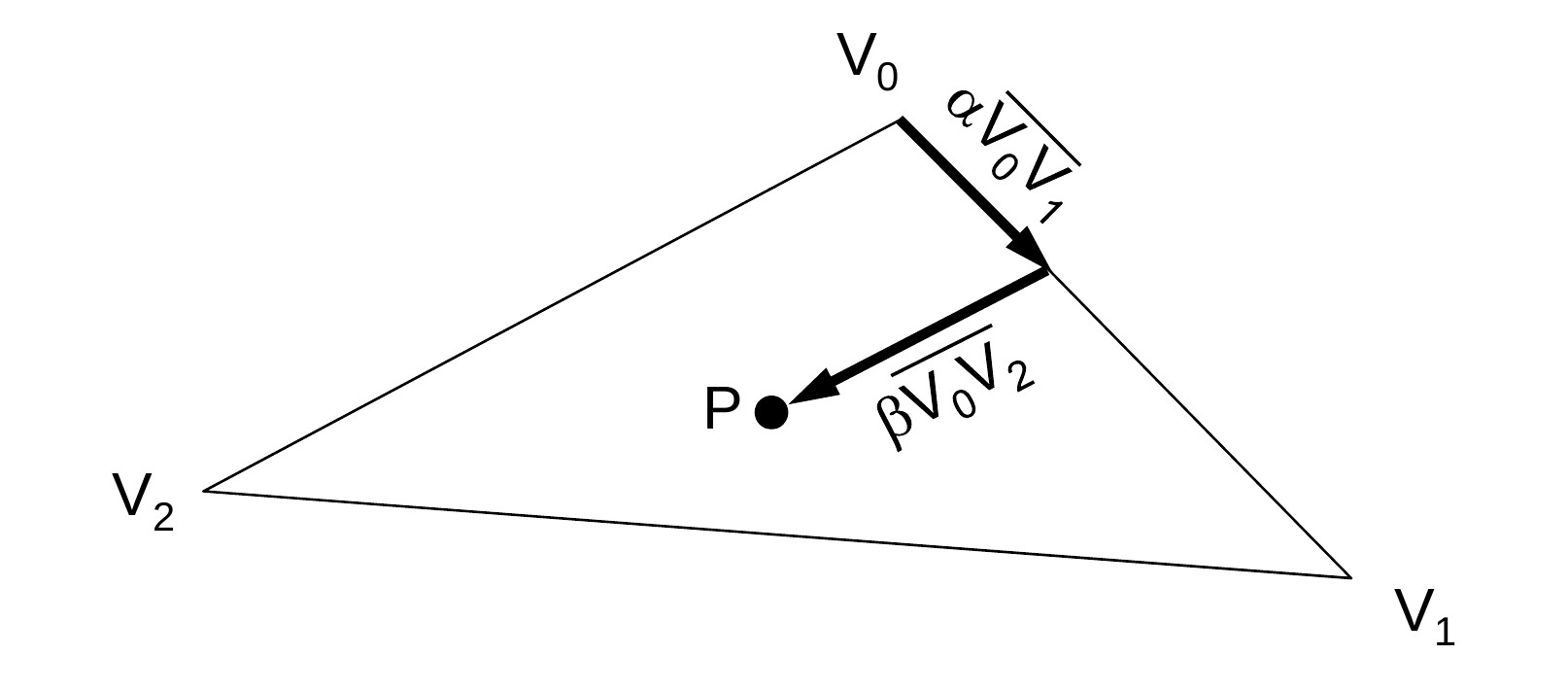}
\caption{Parametric presentation for determining whether the intersection point $P$ of the test particle orbit step with the plane of the wall triangle is inside the triangle.}
\label{fig:parametric_presentation}
\end{center}
\end{figure}

A presentation based on two parameters, $\alpha$ and $\beta$ (see Figure \ref{fig:parametric_presentation}), is used to determine whether the collision with the triangle's plane has occurred inside the boundaries of the triangle. This is true if $\alpha \ge 0$, $\beta \ge 0$ and $\alpha + \beta \le  1$. If several candidate triangles fulfill this criterion, the one with the smallest $t_{i}$ is selected and the wall collision point is returned by the routine.

This algorithm, originally developed for computer graphics applications, has been optimized to use as little CPU time as possible. 
A function $f(\mathbf{r})$ giving a lower limit for the distance to the wall is initialized before the actual simulation starts for checking the possibility of a collision. In addition, the wall elements are arranged into a bounding volume hierarchy tree during the initialization phase and, thus, the volumes that bound the wall elements and collide with the volume bounding the test particle orbit step from $\mathbf{r_{i}}$ to $\mathbf{r_{i+1}}$ can be found in logarithmic search time. This twofold mechanism has proved to be the most efficient the authors are aware of, and requires only a small percentage of the total simulation time.

Although the guiding center formalism offers a method to overcome the high computational demands of following full gyro-orbits, it may lead to false interpretations of, e.g., wall power loads. If the guiding center orbit is close to parallel to the plane of the wall structure, the guiding center may proceed quite far without colliding, even when it is closer than one Larmor radius to the wall elements, whereas the full orbit would collide immediately. Aware of this, we have adopted a mixed scheme: when the particle is no more than one Larmor radius away from the wall structures, the code starts to follow the full gyro-orbit in addition to the guiding center, and if no collision is observed when the particle again recedes farther away from the wall, full orbit-following is abandoned. This approach saves a significant amount of computation time as compared to complete gyro-orbit following, doesn't noticeably increase the computational effort compared to the pure guiding center method, and gives a more accurate estimate of the wall collision location than pure guiding center simulation.

\section{Diagnostic tools}
\label{sec:diagnostics}

During a simulation, test particle data can be gathered into N-dimensional histograms called \emph{distributions}. The distributions can have up to six dimensions, and common to all are \emph{time} and \emph{test particle species}. The remaining 1--4 dimensions are used for the desired phase space coordinates. After each time step $\Delta t$, a weight $W_i$ is added to all active histograms, where $W_i$ is a physical quantity to be diagnosed. 
For compatibility with transport codes, most distributions are produced as 1-D radial profiles in $\rho_p$.

The distributions are updated after each orbit-following time step as follows. First, the locations of the beginning and end point of the time step are determined for each dimension of the distribution. 
If the locations of the beginning and the end of the time step differ, the value $W$ is distributed between the bins crossed, weighed by the fraction of the time step spent in each bin. If $W$ changes during the time step, the value of $W$ is linearly interpolated in time. The exception to this rule is the $j\times B$ torque, as it is gathered after the simulation, as will be described below.

The calculation of the following radial profiles is implemented in ASCOT: particle density~(\ref{eq:density}), energy density~(\ref{eq:energydensity}), parallel energy density~(\ref{eq:parallelenergydensity}), parallel current~(\ref{eq:jpar}), toroidal current~(\ref{eq:jtor}) collisional power deposition to the plasma~(\ref{eq:powerdeposition}), toroidal $j\times B$ torque~(\ref{eq:jxbtorq}), toroidal collisional torque~(\ref{eq:colltorq}), toroidal torque from changes in toroidal canonical momentum $P_\phi$~(\ref{eq:pphitorq}), and particle and energy sources and sinks from CX reactions. The corresponding weights are
\begin{align}
\label{eq:density}
W_n &= w \Delta t \\
\label{eq:energydensity}
W_E &= E w \Delta t \\
\label{eq:parallelenergydensity}
W_{E_\parallel} &= E \xi^2 w \Delta t \\
\label{eq:jpar}
W_{j_\parallel} &= e\mathrm{v}_\parallel w  \Delta t \\
\label{eq:jtor}
W_{j_\phi} &= e \mathrm{v}_\phi w \Delta t \\
\label{eq:powerdeposition}
W_{P,c} &= -\Delta E_d w \\
\label{eq:jxbtorq}
W_{\tau,jxB} &= -e \Delta\psi_p w \\
\label{eq:colltorq}
W_{\tau,c} &= -R \left(\Delta p_{\parallel,d} \frac{B_T}{\left|B\right|}\right) w \Delta t \\
\label{eq:pphitorq}
W_{\tau,p_{\phi}} &= -e \Delta P_\phi w.
\end{align}
Here $w$ is the test particle weight factor indicating how many real particles it represents, $E$ the kinetic energy, $\Delta\psi_p$ and $\Delta P_\phi$ the changes in particle position and canonical toroidal momentum due to the orbit integration, and $\Delta E_d$ and $\Delta p_{\parallel,d}$ are the deterministic changes in energy and parallel momentum due to collisions with the background during $\Delta t$.

 Any distribution can be produced as a function of $\rho_p$, or $(R,z)$. In addition, the density distribution~(\ref{eq:density}) is available in four phase space dimensions $(R,z,\mathrm{v}_\parallel,\mathrm{v}_\perp)$ or $(R,z,\xi,E)$. Distributions depending on interactions with the background, i.e., the power deposition~(\ref{eq:powerdeposition})~and the collisional torque~(\ref{eq:colltorq}), are produced separately for each background species. This is possible because the collisional contributions to the torque and power depositions are calculated directly as moments of the collisional term in the kinetic equation: although the distribution function is represented by the test particles, this does not mean that, e.g., the collisional torque deposition needs to be calculated from the absolute change in the particle's momentum, as is done in ASCOT3. A proper derivation reveals that only the deterministic particle motion contributes to the collisional depositions. This approach, adopted in ASCOT4, also makes it straightforward to divide the collisional contribution between different background species. 

It should be noted that the torques, Eqs.~(\ref{eq:jxbtorq}),~(\ref{eq:colltorq}) and~(\ref{eq:pphitorq}), contain only the component of $\mathbf{\tau}$ parallel to $\mathbf{\hat{z}}$, which is the torque due to a toroidal force. The expression for $j\times B$ torque, Eq.~(\ref{eq:jxbtorq}),  is obtained by calculating $\mathbf{\tau}_{jxB}\cdot\mathbf{\hat{z}} = \left(\mathbf{j}\times\mathbf{B}\right)\times\mathbf{R}\cdot\mathbf{\hat{z}}$, which, assuming an axisymmetric magnetic field $\mathbf{B}=g\nabla\phi+\nabla\psi_p\times\nabla\phi$, reduces to $\mathbf{j}\cdot\nabla\psi_p=q\dot{\psi}_p$. Integrating this expression in time gives the change $\Delta\psi_p$. This result is particularly useful for numerical implementation: it becomes possible to calculate the $j\times B$ torque as a function of $\rho_p$ after the simulation is done, simply finding out which histogram bins belong to the interval defined by the particle's initial and final locations in $\rho_p$.

After the simulation is finished, the distributions are normalized by the volumes of the bins to obtain the density. For radial distributions, the volumes of flux surfaces are obtained by direct integration for closed flux surfaces, and by Monte Carlo integration for open flux surfaces. For distributions in $(R,z)$ space, the volumes of the bins are $dV(R,z) = \pi d(R^2) dz$. The volume elements of velocity space are $dV(\mathrm{v}_\parallel,\mathrm{v}_\perp) = d\mathrm{v}_\parallel d\mathrm{v}_\perp$ and $dV(\xi,E) = d\xi dE$.

In addition to the distributions integrated over the history of the particles, ASCOT also gives the detailed information on the test particles ending up at the wall structures. Not only the location at the wall but also the energy and direction of the particle hitting a material surface are recorded. This information is important when the power and particle fluxes of energetic particles on plasma-facing components have to be evaluated, e.g., to assess that the operational limits of the wall materials are not exceeded.

\section{Benchmark between the new and old versions of ASCOT}
\label{sec:benchmark}
\begin{figure}[!h]
\begin{center}
\includegraphics[width=0.45\textwidth]{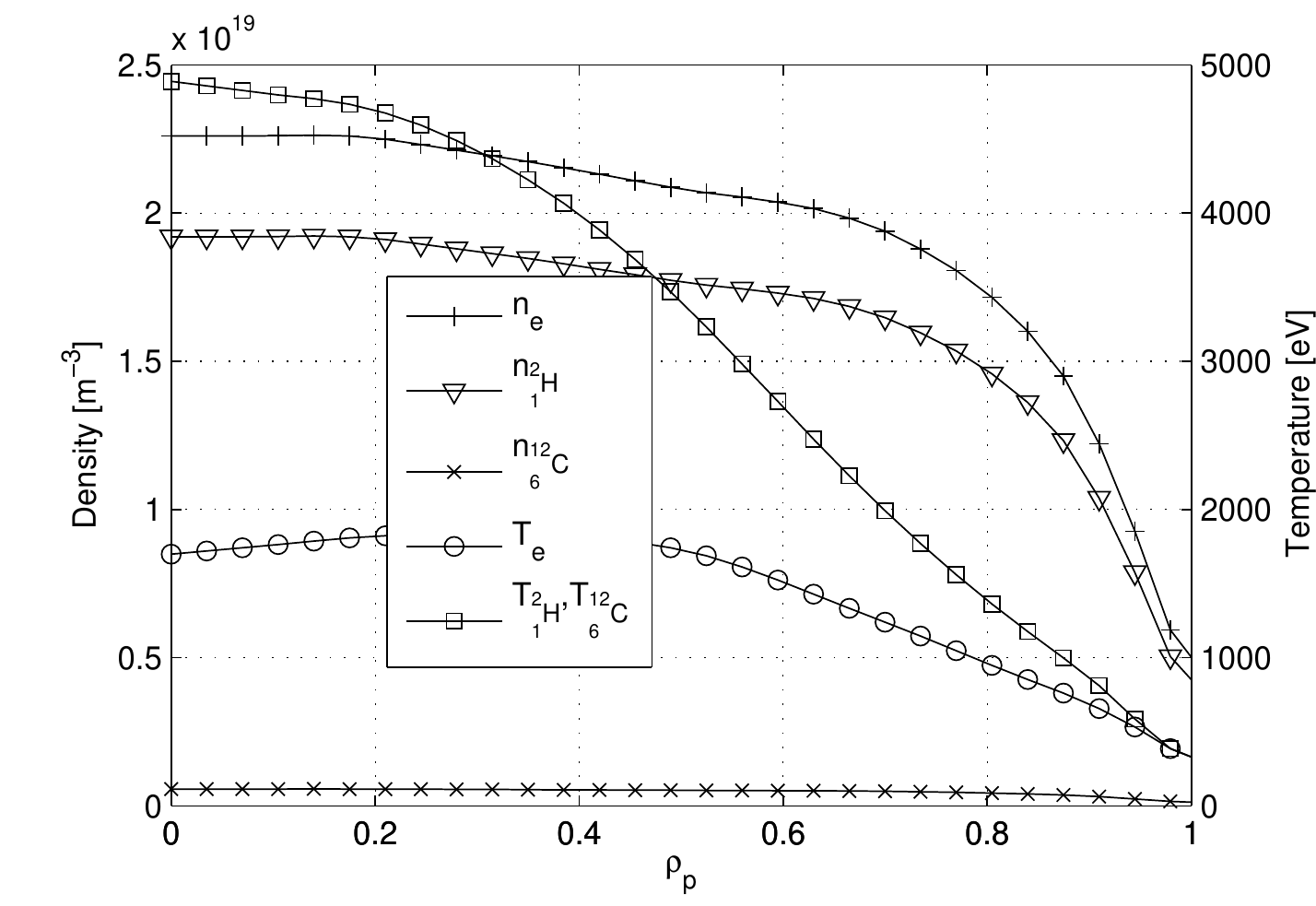}
\caption{Plasma density ($n$) and temperature ($T$) profiles for the benchmark case. The temperatures of the ion species ($^2_1$H and $^{12}_6$C) are assumed equal.}
\label{fig:plasma_density_temperature}
\end{center}
\end{figure}

To carry out a detailed comparison of the results produced by ASCOT3 and ASCOT4, we have chosen an axisymmetric JET-like magnetic background with simple 1D plasma profiles, and carried out a slowing-down simulation of 200~000 test particles representing fast ions from neutral beam injection. The toroidal field strength at the magnetic axis is 3 T and the plasma current is 1.8 MA. During the slowing-down simulation, the test particles are followed with guiding center equations of motion and Coulomb collisions are applied until the particles reach an energy equal to the local background ion temperature. The ion and electron temperatures and densities for the discharge are presented in Fig.~\ref{fig:plasma_density_temperature}. Despite the capability of handling 3D wall data, in this comparison, the wall limiting the calculation regime is taken simply as an axisymmetric structure defined as a contour in the poloidal plane. 

\begin{figure}[!h]
\begin{center}
\includegraphics[width=0.5\textwidth]{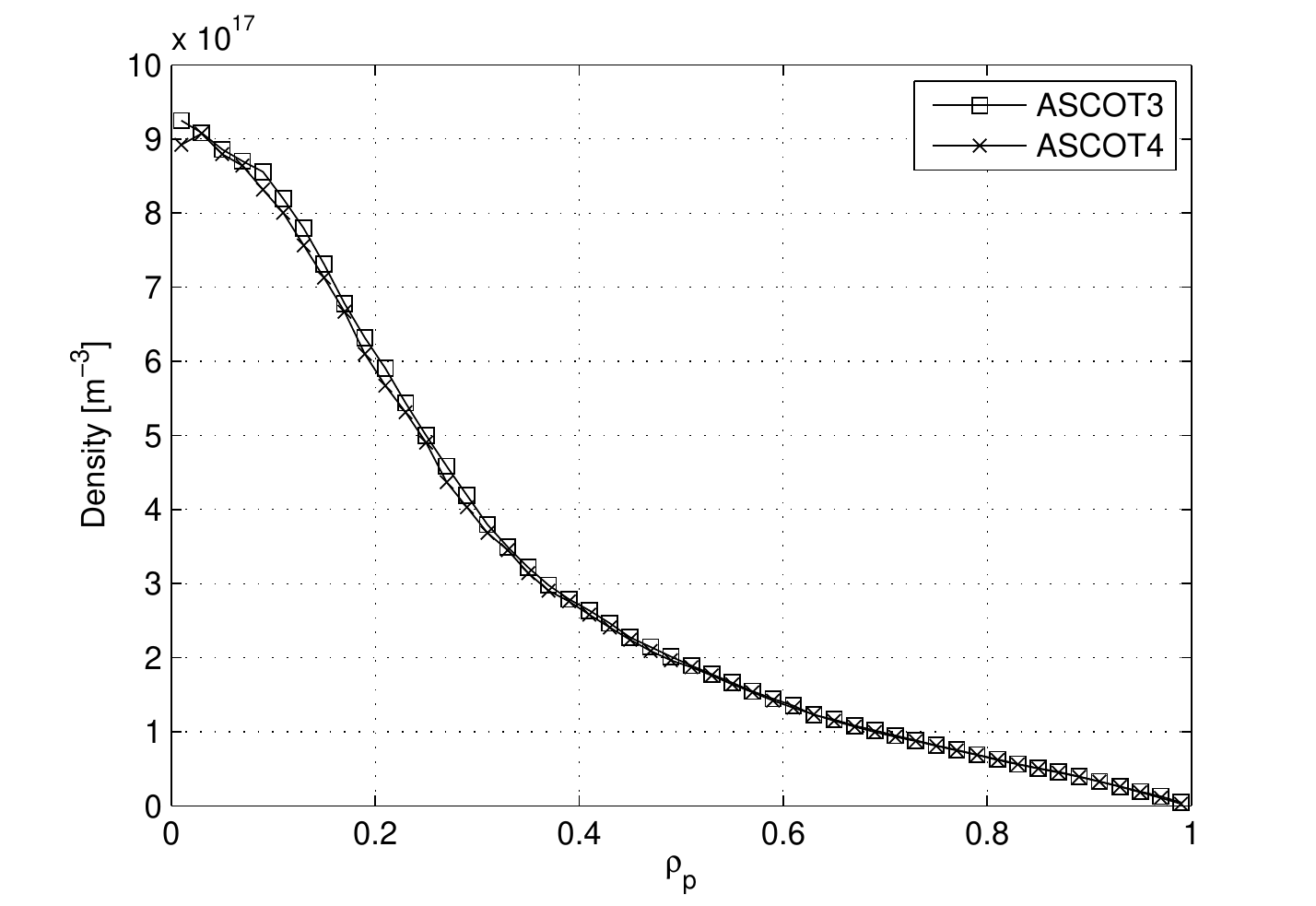}
\caption{The slowing-down density of fast NBI ions calculated with the old and the new version of the code, ASCOT3 and ASCOT4, respectively.}
\label{fig:fast_density}
\end{center}
\end{figure}

\begin{figure}[!h]
\begin{center}
\includegraphics[width=0.5\textwidth]{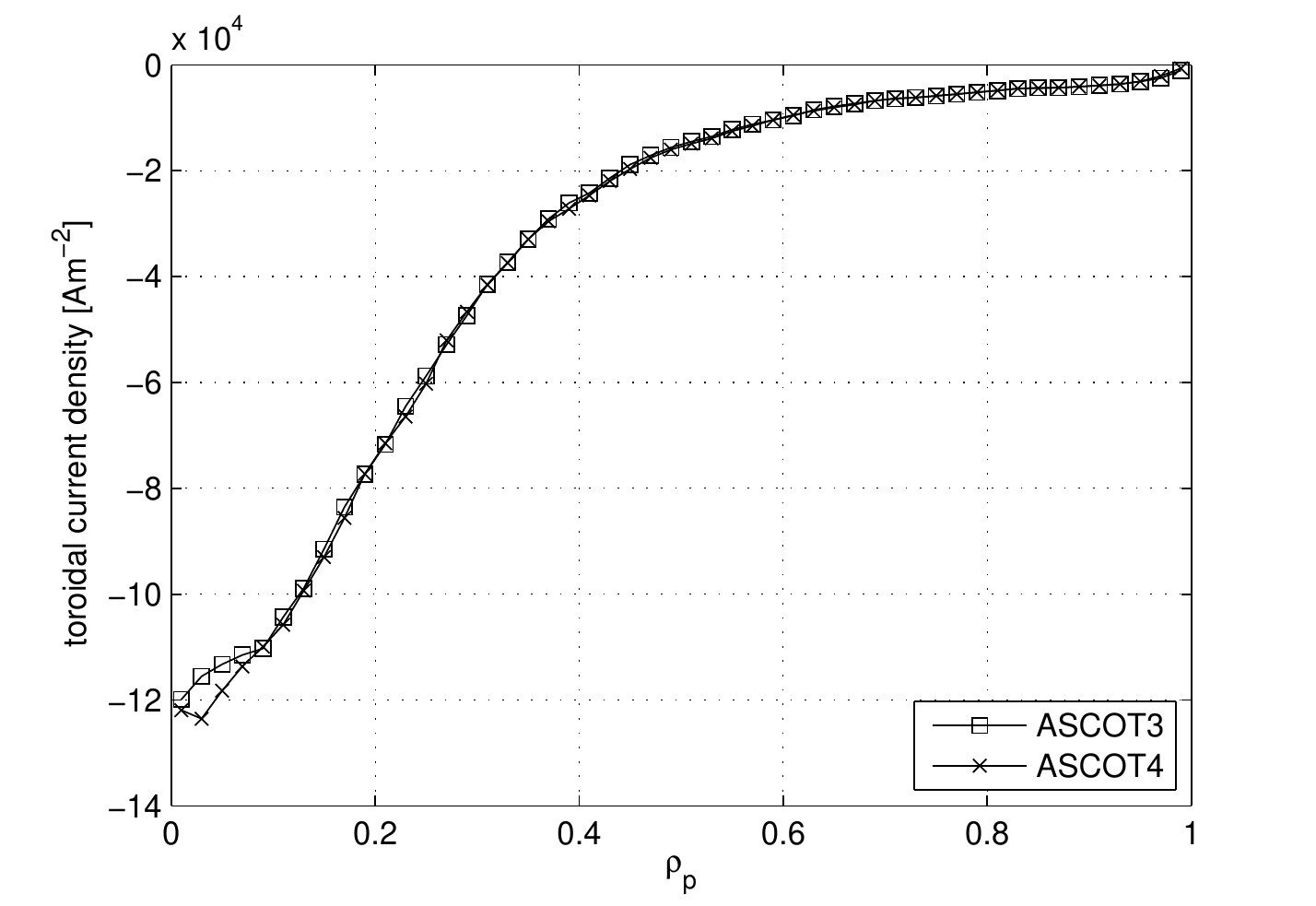}
\caption{The toroidal current density of fast NBI ions calculated with ASCOT3 and ASCOT4.}
\label{fig:toroidal_current}
\end{center}
\end{figure}

\begin{figure}[!h]
\begin{center}
\includegraphics[width=0.5\textwidth]{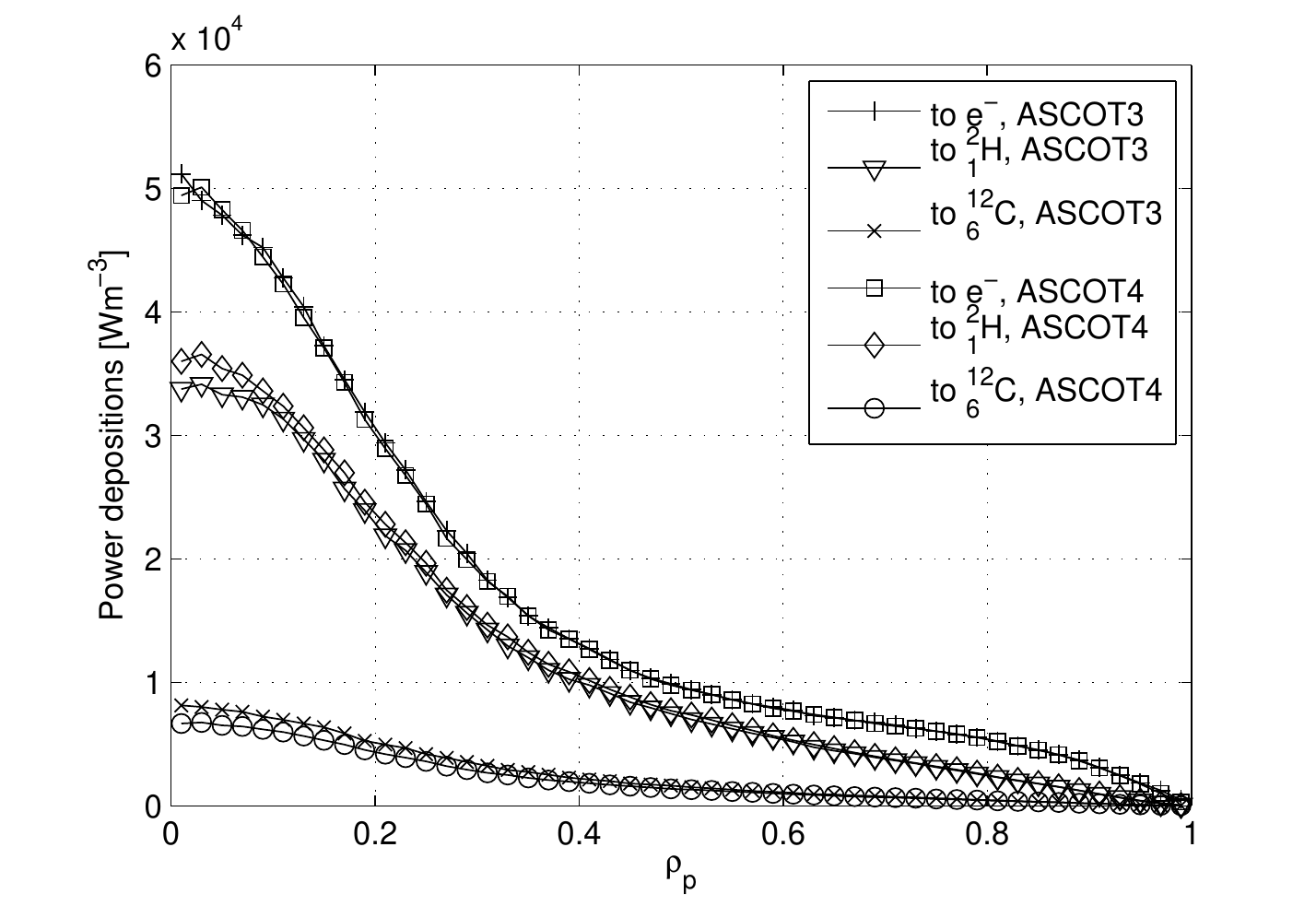}
\caption{Collisional power deposition to different background plasma species due to fast ions from NBI as calculated by ASCOT3 and ASCOT4.}
\label{fig:power_depositions}
\end{center}
\end{figure}

\begin{figure}[!h]
\begin{center}
\includegraphics[width=0.5\textwidth]{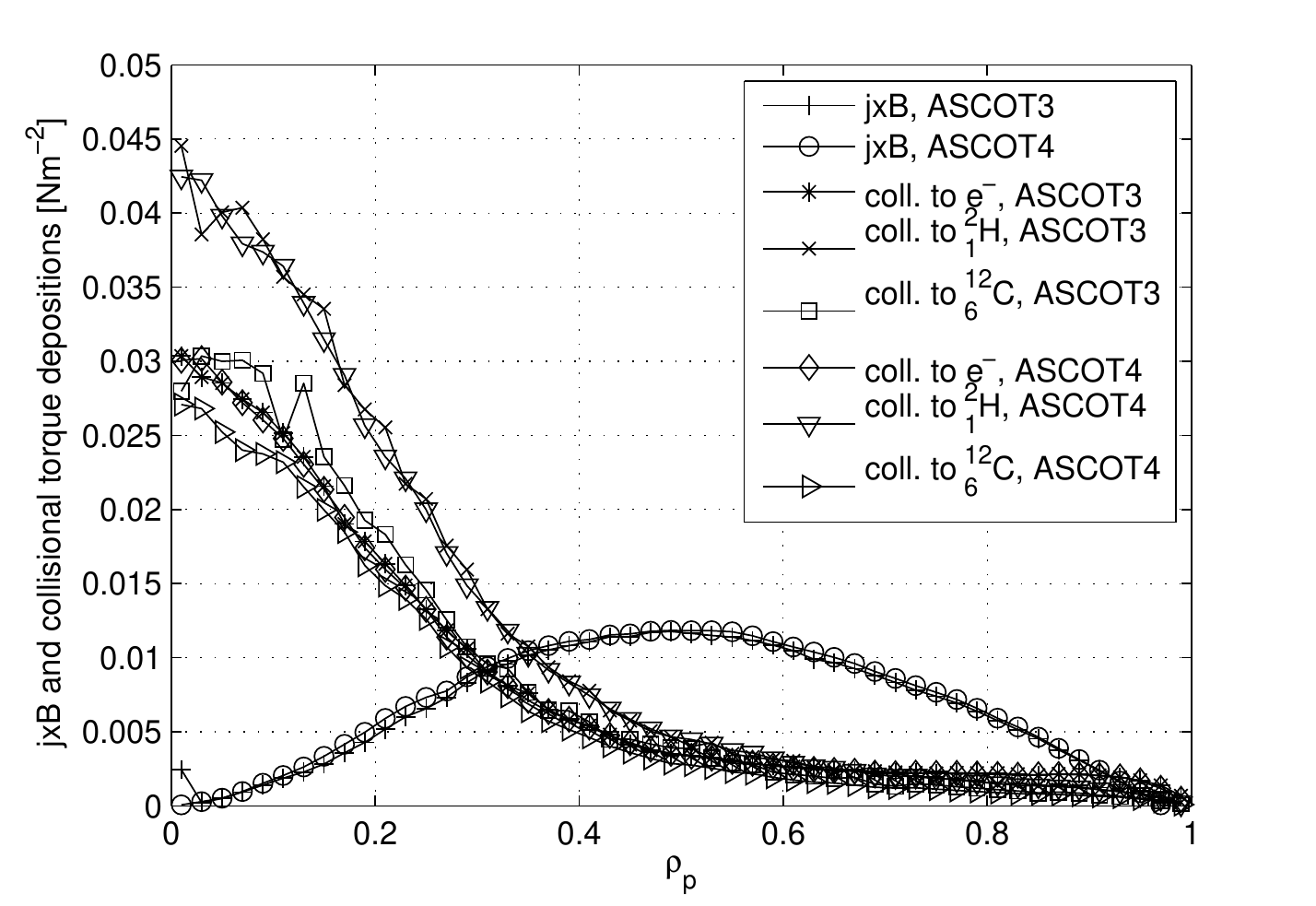}
\caption{The $j\times B$ and the collisional torque deposition from neutral beam ions to the background plasma species. Notice the smoother results provided by ASCOT4.}
\label{fig:torque_depositions}
\end{center}
\end{figure}

As one of ASCOT's applications is to produce source terms for 1D transport codes (being a part of the JINTRAC suite of codes and belonging to the European Transport Solver (ETS) within the ITM framework), the quantities of high interest in our benchmark are the fast ion density, toroidal current density, toroidal $j\times B$ torque, and collisional power and torque depositions, given as a function of the radial coordinate $\rho_p$. These distributions are presented for the current study in Figs.~\ref{fig:fast_density}--\ref{fig:torque_depositions}. The overall agreement of the results is very good. The 1-D fast ion densities and toroidal current densities in Figs.~\ref{fig:fast_density} and~\ref{fig:toroidal_current}, respectively, are practically identical. The power depositions in Fig.~\ref{fig:power_depositions} match well, and the small deviations reflect the different methods for collecting the data: in ASCOT4, the depositions are calculated from the distribution function as described in Sec.~\ref{sec:diagnostics}, and not from the absolute changes in the test particle phase space coordinates, as is done in ASCOT3. The advantages of this approach become evident in Fig.~\ref{fig:torque_depositions}. The collisional torques recorded with ASCOT4 are less noisy than the torques recorded with ASCOT3, and simulations with a smaller number of test particles would highlight this effect. The $j\times B $ torques presented in Fig.~\ref{fig:torque_depositions} agree well between the codes apart from the small anomaly in the plasma core produced by ASCOT3. 

\begin{figure}[!h]
\begin{center}
\includegraphics[width=0.5\textwidth]{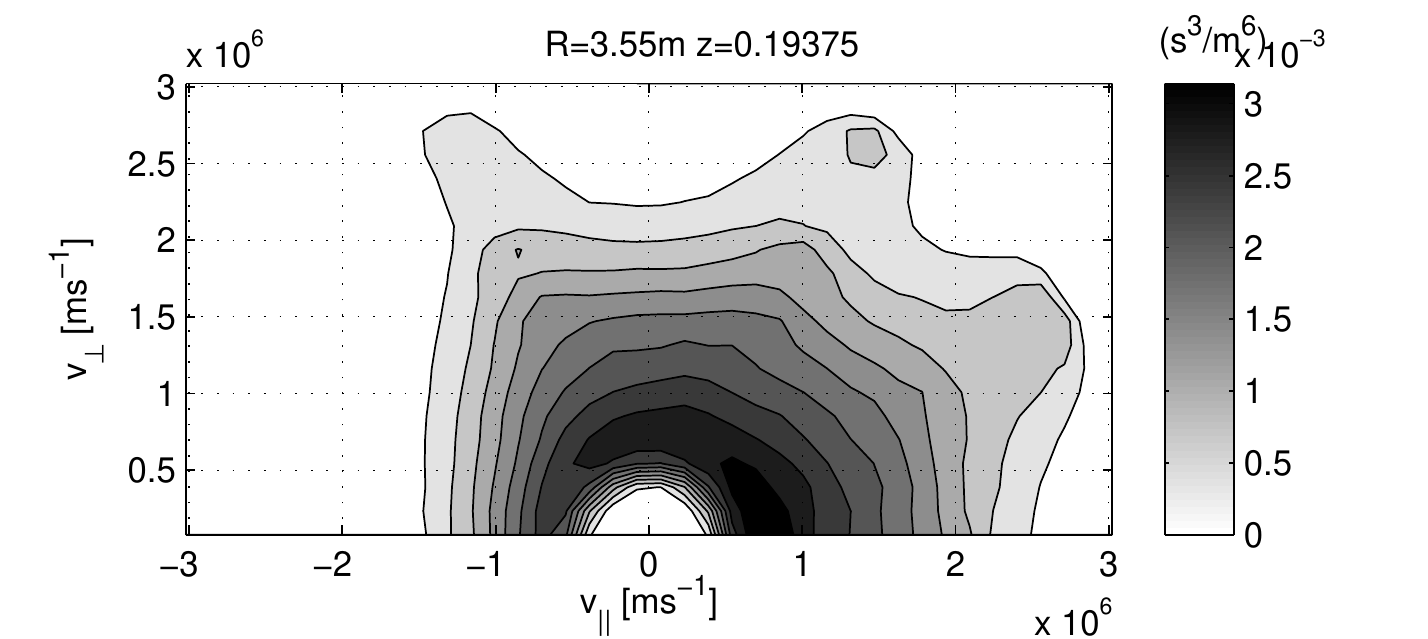}
\caption{Local ($R=3.55,z=0.19375$) NBI slowing-down velocity space distribution from ASCOT4.}
\label{fig:ascot4_4d_density}
\end{center}
\end{figure}

\begin{figure}[!h]
\begin{center}
\includegraphics[width=0.5\textwidth]{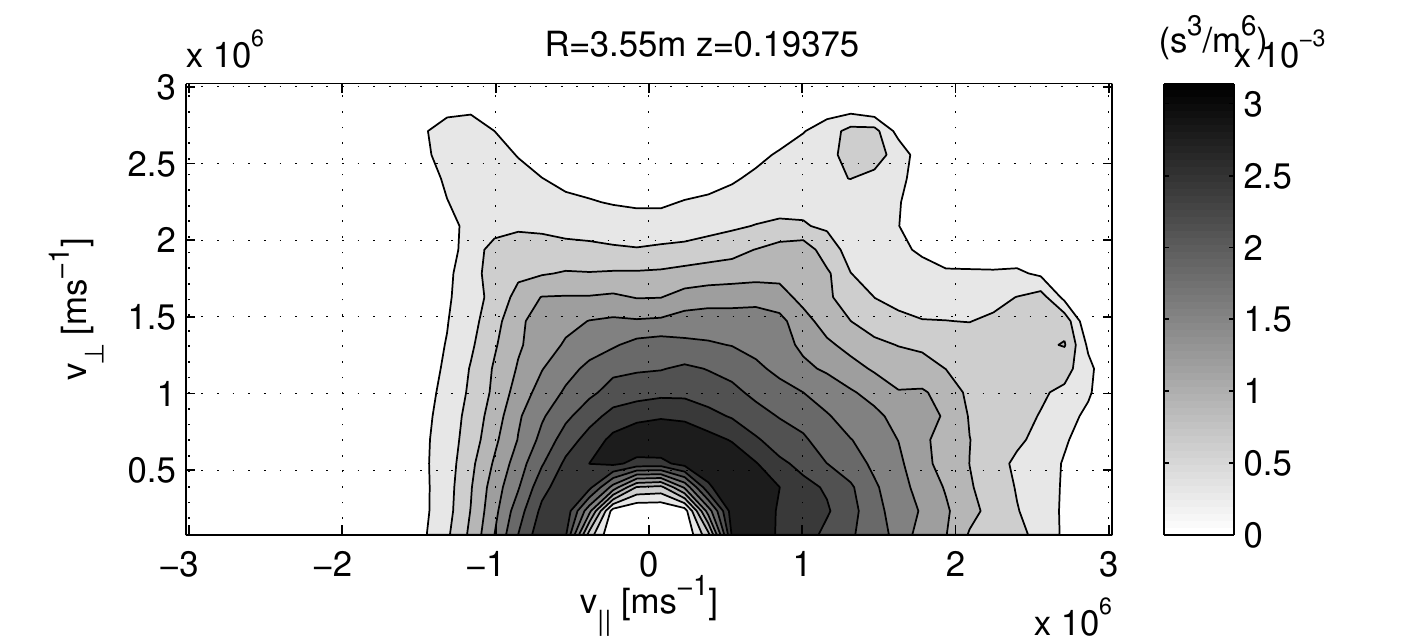}
\caption{Local ($R=3.55,z=0.19375$) NBI slowing-down velocity space distribution from ASCOT3.}
\label{fig:ascot3_4d_density}
\end{center}
\end{figure}

Although only the $\rho_p$ profiles are important if the output of ASCOT is to be used as input for transport codes, a comparison of the 4D density distributions can be considered as a more thorough test: the $\rho_p$ profiles are moments of the actual distribution function, and integrals of two different functions may still agree. The spatially local test particle distribution functions calculated with ASCOT4 and ASCOT3 are presented in Figures~\ref{fig:ascot4_4d_density} and~\ref{fig:ascot3_4d_density}, respectively, for one spatial position, showing very good agreement between the codes. In general, ASCOT4 appears to reliably reproduce the results of ASCOT3 when considering configurations where the less refined methods used in ASCOT3 are not expected to strongly affect the results.

\section{High performance and high throughput computing}
\label{sec:hightech}
The new version of ASCOT is a highly parallel code written in modern FORTRAN. The current record (16th August 2012) is a successful test run with $2^{14}=16384$ parallel processes on the International Fusion Energy Research Centre's Helios supercomputer in Rokkasho, Japan.
\subsection{Data parallel processing}
The problem of following non-interacting particles in a static background is straightforward to parallelize because the calculations can be made independently for each particle. In a typical case, a large number of test particle initial parameters are first stored in a file. Then multiple copies of ASCOT acquire their own slice of the input data and perform the calculations independently. Finally, the distributions as well as other output quantities are merged. There is support for two complementary mechanisms for executing ASCOT in parallel, one for high performance computing (HPC) and another for high throughput computing (HTC). The former uses supercomputers while the latter exploits idling workstations.

The widely used Message Passing Interface~\cite{MPI} (MPI) is the parallelization method of choice when ASCOT is executed on supercomputers. This allows up to thousands of parallel processes for the calculations, while the MPI library provides for the distribution of tasks among the parallel processes. Since file access is often the bottleneck in large parallel jobs, ASCOT uses a single process to read the files from the disk and broadcast and scatter the contents to all the parallel processes using MPI. Similarly, only one process writes the results to the output files.

The ASCOT code is compatible with Condor distributed computing software~\cite{condor-practice}. Condor allows ASCOT users to harvest the idle time of workstations by executing ASCOT on them. All the necessary input files and the ASCOT binary are sent to the remote workstation by Condor, so no common file system is needed. Each ASCOT process produces an output file, all of which are returned to the user's local workstation when the process finishes. When all the processes have been successfully finished, an auxiliary program combines the results into a single output file.

\subsection{External and internal libraries}

At the time of compilation, ASCOT requires only a single external library to be provided by the host system: HDF5~\cite{HDF5}. Practically all ASCOT output and an increasing part of the input is in the HDF5 format. It is a binary format widely supported both in HPC facilities and programming languages. Using HDF5 offers high-performance time-resilient storage for ASCOT data. An extensive set of analysis and visualization tools have been written for MATLAB and are available for the users. 

Inside ASCOT, parts of several large and small libraries are used. The code is stored in a version control system and, within this repository, the code base also includes parts of the following libraries: PSPLINE~\cite{pspline} for splines, Kracken~\cite{kracken} for parsing command line arguments, QUADPACK~\cite{quadpack} for numerical integration, and SLATEC~\cite{slatec} for elliptical integrals. 
The MPI library is usually used with ASCOT, but if it is not available, stubs~\cite{MPISTUBS} are used instead. For random numbers we use the well-known Mersenne Twister algorithm and library~\cite{MatsumotoMersenneTwister}.

\section{Summary and future work}
\label{sec:summary}

With the goal of providing the fusion community with a comprehensive test ion code optimized for fusion applications, we have redesigned the Monte Carlo orbit-following code ASCOT. A formalism that allows solving the kinetic equation for minority ions, be they energetic ions or impurity species, in a manner where the Hamiltonian particle motion and collisions are treated consistently, was developed both for the full gyro motion and in the guiding center formalism. To incorporate all the relevant physics, ASCOT was written to operate in full 3D, thus automatically including all neoclassical drifts. It uses a physical 3D bounding surface, corresponding to the first wall of a fusion device, and includes models for MHD processes relevant for fast ion distribution as well as background flow and atomic reactions that are important for impurity ions. ASCOT features first-principles sources for energetic particles corresponding to alphas and neutrons from thermonuclear, beam-target and beam-beam fusion reactions, as well as ions resulting from neutral beam injection. A simple model for ICRH-accelerated ions is also provided. Simulation results are given as N-dimensional distributions, where up to two velocity dimensions and two spatial dimensions can be assigned. Furthermore, the distribution of ions ending up at the first wall is available, including the information of their energy and direction. To this end, an accurate and efficient description of the first wall and a sophisticated wall-collision algorithm were incorporated into the code.

In the process of redesigning ASCOT, shortcomings in the conventional approach were discovered and fixed. First and foremost was the use of a full gyro motion collision operator when the Hamiltonian motion of test ions was treated within the guiding center formalism. Furthermore, for historical reasons, the collision operator used different velocity space coordinates than those used for the equations of motion. Yet another discrepancy, deriving from the historical development of fast ion codes, is the inconsistency in integrating the Hamiltonian motion with high-level integrators while often treating the deterministic part of the collision operator with the very crude Euler method.

In the new ASCOT, Coulomb collisions are now treated consistently with the Hamiltonian motion and, in the guiding center formalism, spatial diffusion, as given in Eq.~(\ref{eq:spatial}), is introduced in addition to the velocity space diffusion. The remaining question is whether the collisional contribution has to be treated as accurately as the Hamiltonian motion. Work to implement the consistent numerical integration schemes is ongoing.









\end{document}